\documentclass[reprint,prl,  amsmath, amssymb, superscriptaddress]{revtex4-1}
\usepackage{graphicx}
\usepackage{dcolumn}
\usepackage{bm}
\usepackage{siunitx}
\usepackage{eufrak}

\begin{document}
\title{Ionization states for the multi-petawatt laser-QED regime}

\author{I. Ouatu}
\email{iustin.ouatu@physics.ox.ac.uk}
\affiliation{Department of Physics, Atomic and Laser Physics sub-Department, Clarendon Laboratory, University of Oxford, Parks Road, Oxford OX1 3PU, United Kingdom}
\author{B. T. Spiers}
\affiliation{Department of Physics, Atomic and Laser Physics sub-Department, Clarendon Laboratory, University of Oxford, Parks Road, Oxford OX1 3PU, United Kingdom}
\author{R. Aboushelbaya}
\affiliation{Department of Physics, Atomic and Laser Physics sub-Department, Clarendon Laboratory, University of Oxford, Parks Road, Oxford OX1 3PU, United Kingdom}
\author{Q. Feng}
\affiliation{Department of Physics, Atomic and Laser Physics sub-Department, Clarendon Laboratory, University of Oxford, Parks Road, Oxford OX1 3PU, United Kingdom}
\author{M. W. Mayr}
\affiliation{Department of Physics, Atomic and Laser Physics sub-Department, Clarendon Laboratory, University of Oxford, Parks Road, Oxford OX1 3PU, United Kingdom}
\author{R. W. Paddock}
\affiliation{Department of Physics, Atomic and Laser Physics sub-Department, Clarendon Laboratory, University of Oxford, Parks Road, Oxford OX1 3PU, United Kingdom}
\author{R. Timmis}
\affiliation{Department of Physics, Atomic and Laser Physics sub-Department, Clarendon Laboratory, University of Oxford, Parks Road, Oxford OX1 3PU, United Kingdom}
\author{C. Ticos}
\affiliation{Extreme Light Infrastructure - Nuclear Physics (ELI-NP), Horia Hulubei National Institute for Physics and Nuclear Engineering, M\u{a}gurele 077125, Romania}
\author{K. M. Krushelnick}
\affiliation{Center for Ultra-Fast Optics, University of Michigan, Ann Arbor, Michigan USA}
\author{P. A. Norreys}
\affiliation{Department of Physics, Atomic and Laser Physics sub-Department, Clarendon Laboratory, University of Oxford, Parks Road, Oxford OX1 3PU, United Kingdom}
\affiliation{Central Laser Facility, UKRI-STFC Rutherford Appleton Laboratory, Didcot, OX11 0QX, United Kingdom}
\affiliation{John Adams Institute, Denys Wilkinson Building, Oxford OX1 3RH, United Kingdom}

\date{\today}

\begin{abstract}
A paradigm shift in the physics of laser-plasma interactions is approaching with the commissioning of multi-petawatt laser facilities world-wide. Radiation reaction processes will result in the onset of electron-positron pair cascades and, with that, the absorption and partitioning of the incident laser energy, as well as the energy transport throughout the irradiated targets. To accurately quantify these effects, one must know the focused intensity on target in-situ. In this work, a new way of measuring the focused intensity on target is proposed based upon the ionization of Xe gas at low ambient pressure. The field ionization rates from Phys. Rev. A 59, 569 (1999) and from Phys. Rev. A 98, 043407 (2018), where the latter rate has been derived using quantum mechanics, have been implemented for the first time in the particle-in-cell code SMILEI [Comput. Phys. Commun. 222, 351-373 (2018)]. A series of one- and two-dimensional simulations are compared and shown to reproduce the charge states without presenting visible differences when increasing the simulation dimensionality. They provide a new way to accurately verify the intensity on target using in-situ measurements.
   
\end{abstract}

\maketitle

There has been tremendous progress towards the construction and envisioned use of multi-petawatt (PW) class laser facilities  across the globe \cite{ColinDansonFacilitiesReview}. 
The CPA mechanism \cite{CPA_mechanism} allows such extreme powers to be achieved at the ELI pillars \cite{ELIpillars1, ELIpillars2}, currently being commissioned \cite{ELINP_progressreport}. 
Equally powerful lasers will be soon built in the United Kingdom \cite{UK_Laser_VULCAN}, France \cite{Apollon_France}, the United States \cite{ZEUS_Michigan, OPAL_Rochester} and South Korea \cite{CoReLS_SKorea}.
A $100$-PW facility will be built in China \cite{SEL_China} and a sub-exawatt one is currently designed for the Russian Federation \cite{Russia_Laser}. 
All this will lead to focused laser intensities higher than \SI{e23}{\watt\per\cm^2} becoming available. 
Such values of the focused intensity open up new regimes of laser-matter interactions, inaccessible up until now. 
For example, physicists will undertake experiments to determine which theoretical model for radiation reaction applies to the laser-plasma interaction. 
It will also be possible to observe relativistic tunneling, initiate Quantum Electrodynamics (QED) cascades \cite{Artmenko2017}, or generate prolific pair production \cite{Bell_Kirk_PRL2008, Bell_Ridgers_PRL2012} in $\gamma\gamma$ or $e^{-}\gamma$ collisions, to name a few \cite{10PW_enabling_science1}. 
The condition to achieve these milestones is to have a high value of the focused intensity. 
Thus, a reliable method to independently determine the peak intensity in the laser focus is essential in order to investigate these new Physics regimes and can therefore find applications in a wide range of different Physics experiments.

Historically, the focused intensity has been diagnosed by separately measuring the pulse's spatial and temporal parameters at low power and extrapolating to the needed, higher, peak power \cite{LoI_ref1}. 
However, there are long and complex amplification chains in any multi PW-laser facility. 
The wavefront, spectral phase and temporal distortions appearing in these chains cause this method to potentially deliver results drastically different from the real peak intensity achieved in experiments \cite{STCs}. 
There is no easy way to correctly account for the chromatic aberrations and frequency chirps causing those significant distortions, invalidating the conventional methods for ultra-high power laser pulse characterization. 

This letter describes a new method to diagnose the peak intensity, based on the laser-field ionization of a rarefied high-Z noble gas that is self-consistently modelled using a fully relativistic laser-plasma particle-in-cell (PIC) computer simulation code. 
New functionalities have been added to the publicly available SMILEI PIC code \cite{SMILEI}, based upon state-of-the art research on field ionization from \cite{Artmenko2017, Kostyukov_Golovanov_2018, Golovanov2020} and \cite{BM1990} with the method to diagnose peak intensities firstly proposed in \cite{Ciappina2019, Ciappina2020_1, Ciappina2020_2}. 
This letter provides the first evidence that computationally efficient one-dimensional (1D) PIC simulations are accurate enough for the inference of an ultra-high power laser's peak intensity. 
This paves the way to the diagnostic's practical implementation at ELI-NP and other multi-PW laser facilities. 

The model used to describe the field ionization of atoms is commonly determined based on the value of the Keldysh parameter \cite{Keldysh1965}:
\begin{equation}
    \gamma_K = \dfrac{E_K}{E_{l}} = \omega_L\dfrac{\sqrt{2 m_e \hspace{0.6mm} IP_i}}{e \hspace{0.5mm} E_{l}},
\end{equation}
where $m_e$ is the electron's rest mass, $IP_i$ the ionization potential (IP) of the atom, $\omega_L$ the light's frequency, $e$ the fundamental unit of charge and $E_{l}$ the peak electric field of the laser pulse. 
$\gamma_K$ dictates in which regime the ionization takes place: multi-photon (MPI) if $\gamma_K > 1$ or tunnel (TI) if $\gamma_K < 1$. 
For current state-of-the-art laser-matter interaction experiments, where $\gamma_K \ll 1$, the Perelomov, Popov and Terent'Ev (PPT) theory for field ionization rates \cite{Popov2004, PPT1966} is often used. 
Following \cite{Kostyukov_Golovanov_2018}, the PPT static electric field ionization rate is:
\begin{equation}
\begin{split}
w_{lm} &= \omega_a \hspace{0.8mm} \kappa^2 \hspace{1.4mm} C^2_{\kappa l^*} \left(2l+1\right) \left( \dfrac{2}{F} \right)^{2n^*-\vert m \vert -1} \\
&\times \dfrac{(l+ \vert m \vert)!}{2^{\vert m \vert} (\vert m \vert)! (l-\vert m \vert)!} \hspace{1.4mm} exp\left( \dfrac{-2}{3F} \right), \\
C^2_{\kappa l^*} &= \dfrac{2^{2n^*}}{n^* \Gamma(n^* + l^* + 1)\Gamma(n^* - l^*)},
\end{split}
\label{eq:PPT}
\end{equation}
where $l$ and $m$ are the orbital and magnetic quantum numbers of the electron, $\omega_a$ the atomic unit of frequency, $\kappa^2 = IP_i / IP_H$ with $IP_H = m_e e^4 / (2 \hbar^2) \approx$\SI{13.6}{\electronvolt}, $F=E/(\kappa^3 E_a)$ with $E$ the local value of the electric field and $E_a$ the atomic unit of electric field, $n^* = Z / \kappa$, \hspace{0.5mm} $l^* = n^*-1$, with $Z$ the ion charge number, and $\Gamma(x)$ the Gamma function. 
The above rate approaches the ADK rate \cite{ADK1986} when $n^* \gg 1$. 


This PPT rate was already implemented in SMILEI and the PIC code is capable of resolving the fast temporal oscillations of the laser electric field, thus eq. \eqref{eq:PPT} does not need to be averaged over one laser period. 

However, quantum tunnelling would only occur if the potential barrier caused by the combined atomic and laser fields is higher than the initial unperturbed energy level of the electron. If this is not the case, the electron would classically escape above the potential barrier in a process called barrier-suppression ionization (BSI) \cite{Donna_BSIdiscovery_1990}. 
BSI happens for $E \gtrsim E_{cr} = E_a \hspace{0.8mm} \kappa^4 / (16Z)$. 
For $E>E_{cr}$ the tunnel rates strongly deviate from numerically calculated rates \cite{Tong_Lin_2005}, as expected. 
A rate correctly describing the transition region between TI and BSI is the Bauer-Mulser (BM) rate $w_{BM} = 2.4 \hspace{0.5mm} \omega_a \left(E/E_a \right)^2 \left( I_H/I_i \right)^2$, proven to work for $E \sim E_{cr}$ \cite{BM1990}. 
However, $w_{BM}$ deviates from numerically calculated rates as $E \gg E_{cr}$ \cite{BM1990, Kostyukov_Golovanov_2018}. 
Derived both classically and quantum mechanically (in the motionless and free electron approximation), the rate dependence on $E$ for $E \gg E_{cr}$ is linear \cite{Kostyukov_Golovanov_2018}: $w_{BSI} = p \hspace{0.9mm} \omega_a (E/E_{a}) \sqrt{(I_H/I_i)}$. 
The coefficient $p$ ranges from $0.62$ to $0.87$, depending on the potential used in the quantum derivation: for the 3D Coulomb potential $p \approx 0.8$. 
Numerically integrating the 1D time-dependent Schr\"{o}dinger equation shows that the two approximations used in the quantum derivation are indeed applicable \cite{Kostyukov_Golovanov_2018}. 
$w_{lm}$, $w_{BM}$ and $w_{BSI}$ are grouped into a piecewise-defined ionization rate \cite{Golovanov2020} which describes the whole range of possible $E$ values at the location of the atom:
\begin{equation}
w(E) \approx \begin{cases}
   w_{lm} \hspace{0.5mm} , & E \leq E_1 \\
   w_{BM} \hspace{0.5mm} , & E_1 < E \leq E_2 \hspace{2.8mm} , \\   
   w_{BSI} \hspace{0.5mm} , & E > E_2
   \end{cases}  
\label{eq:3piecerate}
\end{equation} where $E_1$ and $E_2$ are found by imposing continuity on $w$ at the transition points.
We have implemented this piecewise-defined rate as a new functionality in SMILEI.

The peak intensity diagnostic from \cite{Ciappina2019, Ciappina2020_1} is based on counting the ions of various charge states of a noble gas entering particle detectors.
Comparing the theoretically expected numbers of ions with the experimental results allows inferring the peak intensity \cite{Ciappina2020_2}.

For a fixed value of the laser's normalized vector potential $a_0 = e E_{l} / m_e \omega_{L} c $, a computationally efficient 1D PIC simulation shows how the proportion of the different charge states evolves during the passage of the laser pulse.
Running many such 1D simulations, covering the range $a_0 \in [6.8, ..., 2162.8]$ corresponding roughly to $I \in [10^{20}, ..., 10^{25}] \hspace{0.6mm} \si{\peta\watt}$ for a plane wave, the ionization behavior as the intensity is varied can be obtained.
We ran $15,000$ 1D simulations to finely sample this intensity range.
Xenon (Xe), atomic number $Z=54$, was used, and the initially neutral gas density was chosen such that the electron number density for the fully ionized gas, $n_e$, is low enough to make both collective effects and relativistic self-focusing negligible.
The first condition led to us setting the plasma frequency $\omega_{p} = \sqrt{( n_e \hspace{0.5mm} e^2 / m_e \hspace{0.5mm} \epsilon_0)}$ to be such that the plasma oscillation period is significantly longer than the duration of the laser pulse (which is of O(\SI{10}{\fs}) at ultra-high power laser facilities).
The second condition led to us setting $n_e$ to be such that $P_{L} =\SI{10}{\peta\watt}$ is lower than $P_{cr} = 17\left( \omega_{L} / \omega_{p} \right) ^2\si{\giga\watt}$.
To meet both constraints, we chose the neutral Xenon number density $n_0 = \SI{1.98e12}{\cm^{-3}}$.

\begin{figure}
\includegraphics[width=1.0\columnwidth, height=0.3\textheight]{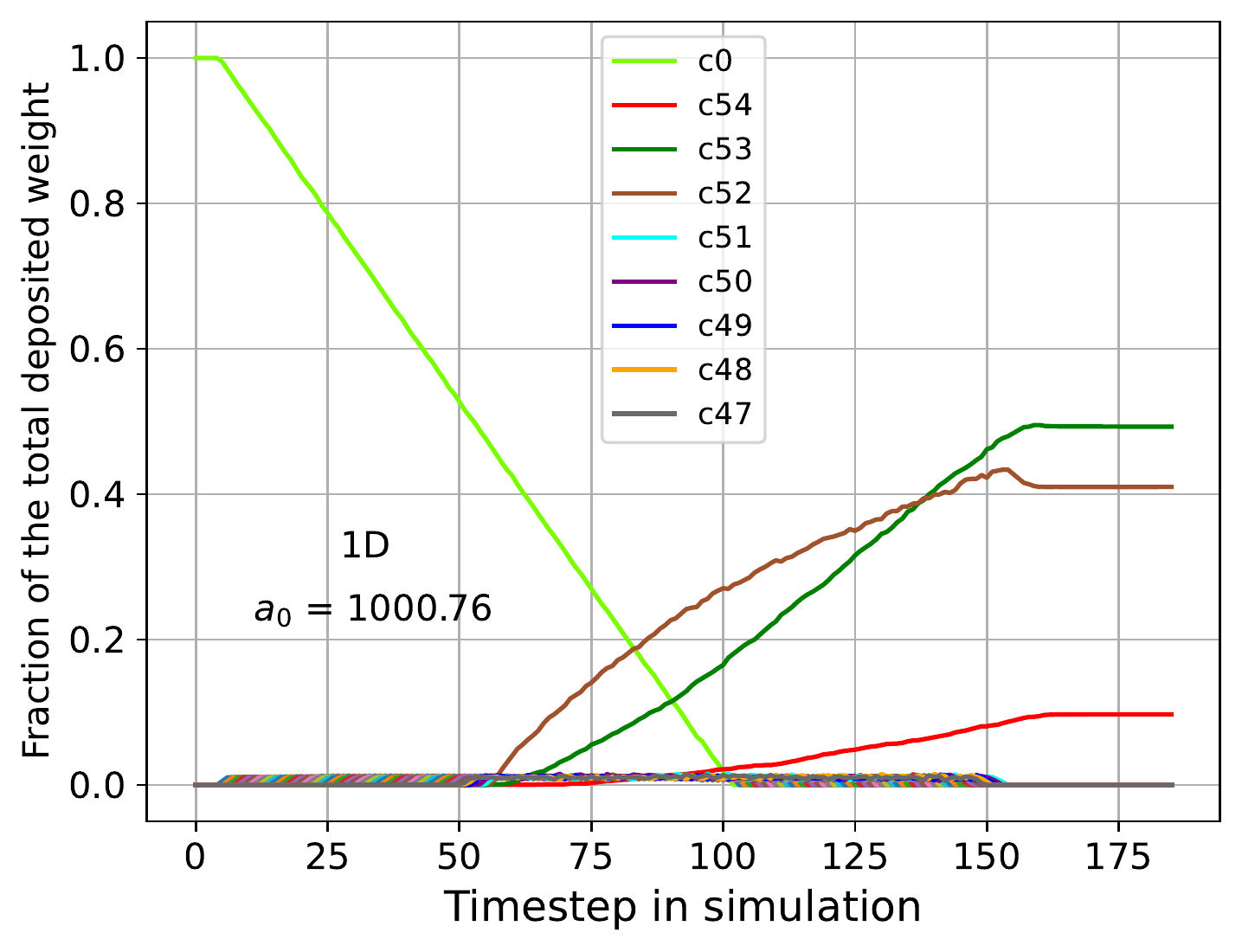}
\caption{Xenon charge states evolution during a 1D PIC simulation. The initial gas number density was $n_0$ = 1.98 $\ \times$ 10$^{12}$cm$^{-3}$ and the laser's normalized vector potential was $a_0$ = 1000.76. The electrons species resulted from field ionization had the Radiation Reaction module turned on. The gas density was low enough to allow neglecting collective plasma effects and relativistic self focusing. \label{fig:Xenon_chargestates_in_time_RRon}}
\end{figure}

Typical results of a one-dimensional (1D) simulation are shown in Fig. \ref{fig:Xenon_chargestates_in_time_RRon}.
The charge state dynamics are displayed as a function of time during the simulation.
The y-axis represents the deposited weights from a particle binning diagnostic for different ion charge states. The electrons resulting from ionization were injected into the simulation as a separate species which had its Radiation Reaction module turned on.
The input laser had a wavelength $\lambda_{L} = \SI{800}{\nano\meter}$ and a Gaussian temporal profile with a full-width at half-maximum (FWHM) of $6$ optical cycles. The extent of the laser's temporal profile was limited to $10$ optical cycles.
The simulation box was $12$ $\lambda_L$ long and the spatial resolution was $8$. On both sides of the box, Silver-M\"{u}ller boundary conditions were chosen for the fields and periodic boundary conditions were chosen for the particles. 
The simulations lasted for $22$ optical cycles and the number of particles per cell (ppc) for the Xe species was $32$.   
To check the consistency of the results obtained relative to the number of ppc, we ran the same simulations with $128$ ppc and obtained the same results as for those with $32$ ppc.

\begin{figure}
\includegraphics[width=1.0\columnwidth, height=0.3\textheight]{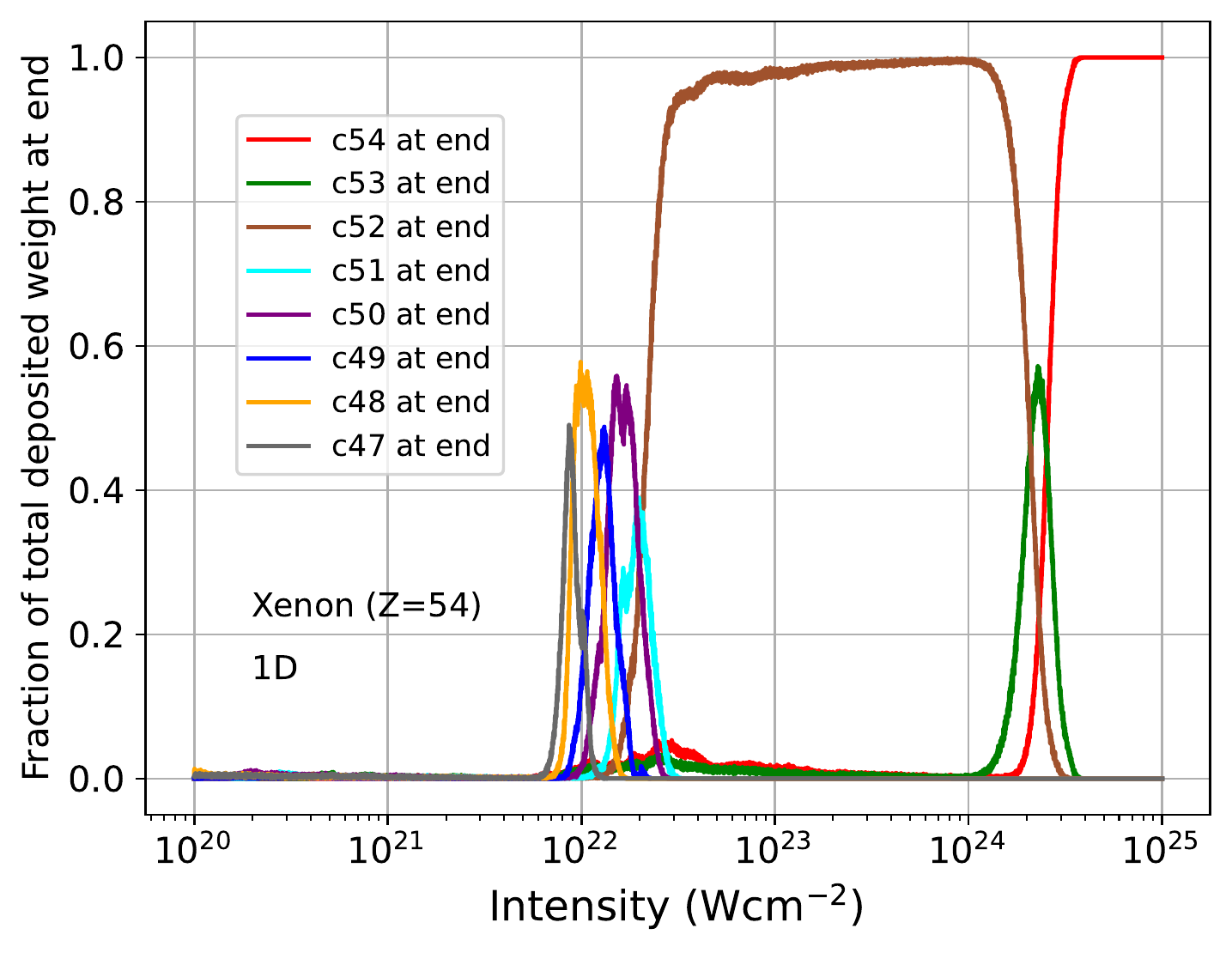}
\caption{Xenon representative charge states' populations at the end of 15,000 1D PIC simulations. 1 datapoint from each curve corresponds to the fraction of the total deposited weight by all charge states, at the end of one such 1D simulation. These $c_n(I)$ curves can be used to calculate the number of ions produced in the focus of a Gaussian beam.  \label{fig:Xenon_1Dchargestates_at_end_RRon}}
\end{figure}

The results for the whole range of intensities appear in Fig. \ref{fig:Xenon_1Dchargestates_at_end_RRon}. 
Each data point from one $c_n$ curve corresponds to the fraction of the deposited weight of the charge state Xe$^{n+}$ relative to the total deposited weights of all states at the end of one simulation at a particular input intensity. 
The charge states values at the end of each simulation are thus plotted against the corresponding intensities.

\begin{figure*}
\includegraphics[width=\textwidth, height=0.30\textheight]{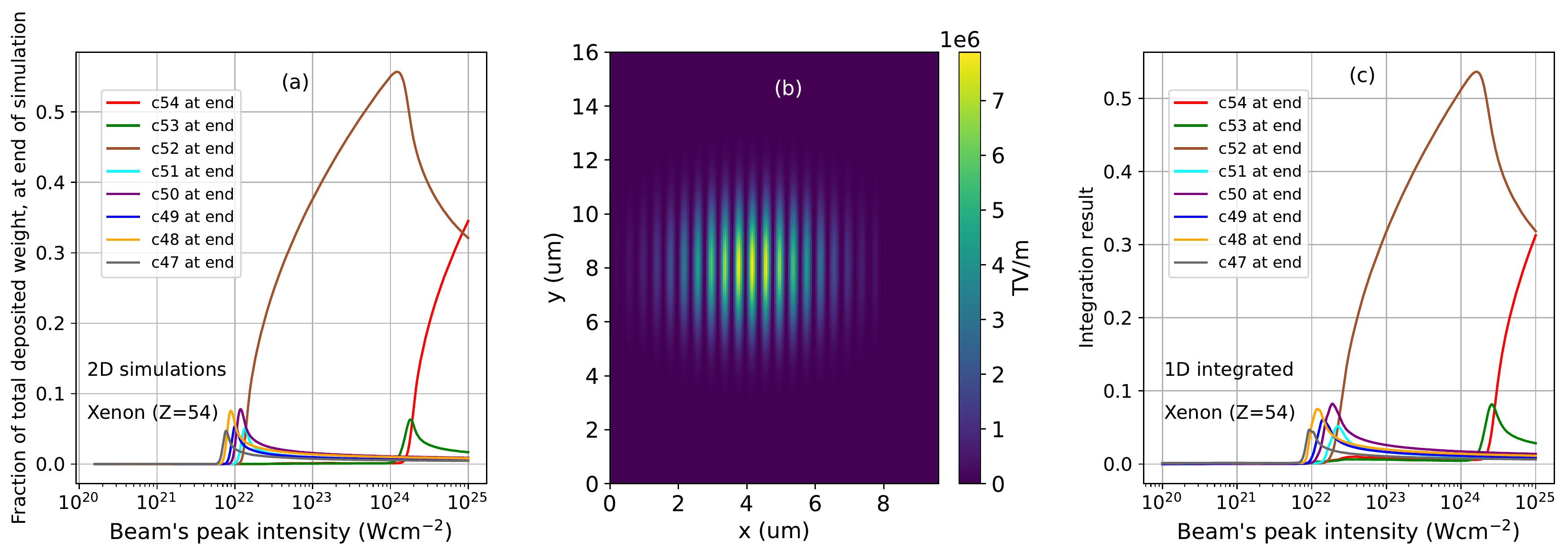}
\caption{(a): Xenon representative charge states' populations at the end of 1,000 2D gaussian beam PIC simulations. 1 datapoint from each curve corresponds to the fraction of the total deposited weight by all charge states, at the end of one such 2D simulation. (b): squared electric-field spatial distribution at one timestep, across the 2D PIC simulation box, for $a_0 = 707.19$ (1 representative example out of the 1,000 runs). (c): Results of integration of the 1D charge states across the 2D PIC simulation's box, showing that there are no important differences relative to (a).
\label{fig:2DGaussian1000simsPure_GaussianBeamPicture_1Dintegrationto2Dresults}}
\end{figure*}

To diagnose the peak intensity by means of these simulations, the numerically-estimated functions $c_n(I)$'s shown in Fig. \ref{fig:Xenon_1Dchargestates_at_end_RRon} are used to infer the number of Xe ions expected to be created at the focus of the laser pulse. 
First, a suitable model for the intensity's spatial distribution $I\left( \vec{r} \hspace{0.5mm} \right)$ is chosen, then the number of Xe ions of charge state $n+$ is obtained from an integral over the focal volume:
\begin{equation}
N \left( Xe^{\hspace{0.5mm}n+} \right) = n_0 \iiint c_n \left( I\left( \vec{r} \hspace{0.7mm} \right) \right) d^3 r .
\label{eq:Integral_cn_s_across_focal_volume_def}
\end{equation}
This gives an accurate estimate for the number of ions created in a real-world experiment.

We have calculated the above with $I \left( \vec{r} \hspace{0.5mm}  \right)$ being a gaussian distribution:
\begin{equation}
I (\vec{r}) = \dfrac{I_m}{1 + \dfrac{z^2}{z_R^2}} \hspace{1.5mm} \exp \left( -\dfrac{2 \rho^2}{w_z^2} \right) ,
\label{eq:Gaussian_Beam}
\end{equation}
with $\rho = \sqrt{x^2 + y^2}$, $I_m$ the peak intensity, $z_R = \pi w_0^2 / \lambda_L$ the Rayleigh range, $w_0 = \SI{3}{\micro\meter}$ the beam waist and $w_z = w_0 \sqrt{1 + (z^2/z_R^2)}$. 
Other descriptions of the focus can be employed \cite{Ciappina2020_2}.

To numerically calculate the integral from eq. \eqref{eq:Integral_cn_s_across_focal_volume_def}, linear 1D interpolators are used to obtain the values of $c_n$ at peak intensities not simulated using the PIC code but required by the integration routine.
For intensity values outside the PIC-simulated range of $[10^{20}, ... , 10^{25}] \hspace{0.7mm} \si{\watt\per\cm^2}$, manual extrapolation for the highly charged states (i.e. $c_{47}$ to $c_{54}$ ones) is done. 
From physical considerations (confirmed by simulated results shown in Fig. \ref{fig:Xenon_1Dchargestates_at_end_RRon}), any highly charged state is not excited for $I < \SI{e20}{\watt\per\cm^2}$. Additionally, only the fully ionized state remains for $I > \SI{e25}{\watt\per\cm^2}$.

Then a midpoint rule in 2 dimensions ($z$ and $\rho$) is employed.
Gaussian beams are symmetric with respect to the azimuthal dimension $\phi$, thus a 2D cylindrical integral is performed, being then multiplied by $2\pi$ to calculate the full 3D integral.
The integration ranges are obtained by considering that below a threshold intensity the $c_{47}$ to $c_{54}$ charge states are not excited. From Fig. \ref{fig:Xenon_1Dchargestates_at_end_RRon}, $I_{thr} = \SI{e20}{\watt\per\cm^2}$.

The integral in $z$ is performed across $[ -z_{max}, z_{max}]$, with
\begin{equation}
z_{max} = z_R\sqrt{\dfrac{I_m}{I_{thr}} -1}.
\end{equation}
At each step along $z$, the integration along $\rho$ is performed across $\left[ 0, \rho_{max} \right]$, where $\rho_{max}$ is obtained by imposing that the intensity at that particular $z$ and $\rho_{max}$ equals $I_{thr}$, so 
\begin{equation}
\rho_{max} = \sqrt{ \dfrac{-w_z^2}{2} \ln\left( \left(1 + \dfrac{z^2}{z_R^2}\right) \dfrac{I_{thr}}{I_m} \right)}.
\end{equation}
The number of steps in $z$ was $N_z = 2^{12}$ and the number of steps in $\rho$, for each $z$ value, was $N_\rho = 2^{12}$.

To verify that the reduced dimensionality of the simulations was not impacting the obtained results, we have also investigated the relationship between intensity and charge state distribution in two-dimensional (2D) simulations.
For an \SI{800}{\nano\meter} beam, with a Gaussian transverse mode, a \SI{3}{\micro\meter} radius at focus, and the same temporal profile as above, the charge states at the end of the simulation for 1,000 different $a_0$ values are presented in the left panel of Fig. \ref{fig:2DGaussian1000simsPure_GaussianBeamPicture_1Dintegrationto2Dresults}. 
The middle panel of Fig. \ref{fig:2DGaussian1000simsPure_GaussianBeamPicture_1Dintegrationto2Dresults} shows the spatial distribution of a representative beam's squared electric field from one such 2D simulation.
The simulation box was 12 laser wavelengths along x and 20 laser wavelengths along y, with SilverM\"{u}ller boundary conditions along x and Periodic boundary conditions along y.

To check the consistency of the results between the 1D and 2D simulations, we performed a 2D version of the 3D integral from eq. \eqref{eq:Integral_cn_s_across_focal_volume_def} of the $c_n$ found in the 1D simulations over a grid of the same size as the 2D simulation box. The right panel of Fig. \ref{fig:2DGaussian1000simsPure_GaussianBeamPicture_1Dintegrationto2Dresults} confirms that the charge states at the end of a 2D simulation reproduce the 1D results: the reduced dimensionality of the simulations is not an impediment to estimating the predicted charge states.

\begin{figure}
\includegraphics[width=1.\columnwidth, height=0.3\textheight]{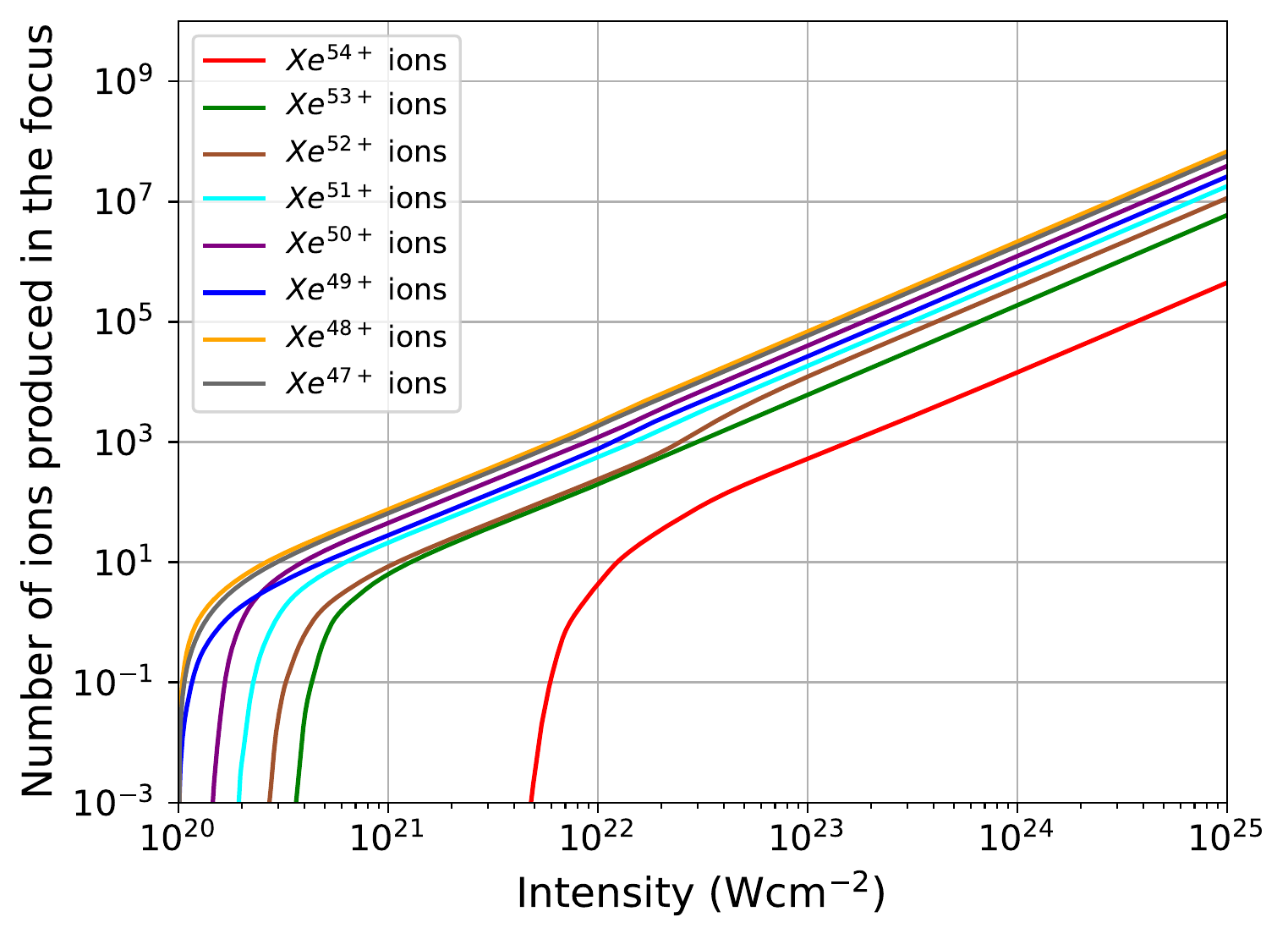}
\caption{Number of Xenon ions produced in the focus of the gaussian beam from eq. \eqref{eq:Gaussian_Beam}. 
One datapoint on each curve corresponds to the number of ions arising from one, fixed, peak intensity $I_m$ value in eq. \eqref{eq:Gaussian_Beam}. 
Results on y-axis are computer calculations for $N \left( Xe^{\hspace{0.5mm}n+} \right)$ from eq. \eqref{eq:Integral_cn_s_across_focal_volume_def}. The integral result was obtained using a 2D Midpoint rule and is denoted by $I_M$ with the number of points used along the two directions $M_{z} = M_{\rho} = 2^{12}$. 
The errors estimated as $\vert \left( I_M - I_{M/2} \right) \vert$, where $I_{M/2}$ represents the integral result from an integration with $M_{z}/2 = M_{\rho}/2 = 2^{11}$ points along the two dimensions, after being multiplied by $n_0$, are too small to be visible on the graph. \label{fig:Integration_Xe}}
\end{figure}

Computational methods are not exact in practice and this could lead to difficulties when comparing to experimental results. One needs to estimate the numerical errors to ensure the veracity of the results.
The $N \left( Xe^{\hspace{0.5mm}n+} \right)$'s from eq. \eqref{eq:Integral_cn_s_across_focal_volume_def}, obtained using the 2D midpoint rule, are presented in Fig. \ref{fig:Integration_Xe}.
There, integral estimates in the theory of converging sequences of numerical approximations \cite{ElemNumAnalysisIowa} are shown (after being multiplied by $n_0$).
That is, the result estimate for one integral, $I$, is a laptop's calculation with the highest number of points it can accommodate, $I_M$, where $M = 2^{12}$ for both $z$ and $\rho$ axes.
The same integral, with half the number of points along both $z$ and $\rho$, was performed and stored as $I_{M/2}$.
The numerical error on a result $I_M$ was determined to be $\vert I_{M} - I_{M/2} \vert$.
The numerical errors approximated in this way, after being multiplied by $n_0$, are too small to be visible when plotted on the figure.

To count the ions from the focus of the beam producing the species presented in Fig. \ref{fig:Integration_Xe}), filtering of the particles based on their charge-to-mass ratio is needed and this is normally done using a Thomson spectrometer. An O(10PW) laser facility is expected to deliver a focused intensity of around $10^{23} \si{\watt\per\cm^2}$.
At least $N_{min} = 50$ ions of a given charge state need to be collected by the spectrometer for a clear differentiation of that charge state from any of the neighbouring parabolas. One therefore needs to calculate the minimum initial gas density to allow $N_{min}$ ions to be collected, taking into account realistic distances from the spectrometer pinhole to the interaction point.

First, the ions are assumed to expand isotropically from the focus (confirmed by the 2D PIC simulations).
Then, a circular aperture of diameter $D = \SI{500}{\micro\meter}$ gives a solid angle of $\Omega = (\pi D^2/4) / R^2$  (the planar area of the aperture is used because its dimensions need to be at least one order of magnitude smaller than the source-aperture distance $R$ based on experimental considerations).
From Fig. \ref{fig:Integration_Xe} the total number of Xe ions produced in the focus is read for any charge state and is denoted as $N^{n+}_{total}$ for charge state $n+$. 
$N^{n+}_{total}$ corresponds to the full $4 \pi$ sr for an isotropic ejection of the ions. 

Only $N_{Th} = N^{n+}_{total} \hspace{1.3mm} \Omega / (4\pi)$ ions will be actually collected by the pinhole, for any $n+$.
For 50 ions of each charge state to pass through the pinhole, the aperture has to be placed at around or less than $R = \SI{0.1}{\cm}$ from the beam's focus.
Clearly, this placement at such a small distance from the source is not realisable in practice. One needs to increase the interaction point - pinhole distance.
Doing so, multiple shots are needed to collect enough statistics for a good differentiation of the parabolas.

Increasing $D$ decreases the spectrometer's capability to discriminate between different charge states and thus while providing a higher ion flux at the detector, a very large diameter aperture is not suitable for the envisioned future experiments.

The solution for a single-shot experiment is to increase the neutral gas number density $n_0$. 
However, one needs to avoid relativistic self-focusing: it would provide an artificially high estimate of the peak intensity. The results presented so far are for simulations with $n_0 = \SI{1.98e12}{\cm^{-3}}$ for Xe. These were repeated for $n_0$ values up to three orders of magnitude higher. No self-focusing effects were observed in the 2D simulations: the numerical results were essentially the same. Increasing $n_0$ by a factor of $1000$ increases the number of ions produced in the focus by the same factor (eq. \ref{eq:Integral_cn_s_across_focal_volume_def}) and therefore a single shot measurement of the peak intensity using Xe is more easily achieved. Depending on the experimental constraints of each facility, one can run additional PIC simulations to check when relativistic self-focusing effects appear in order to choose the most suitable gas density for the optimum source - pinhole distance.

Simulations for another noble gas were performed in the same manner. Argon (Z=18) is completely ionized at $\sim$ $I^{Ar}_{max} = \SI{3e21}{\watt\per\cm^2}$ according to the 1D PIC simulations. For facilities capable of delivering ultra-high power ($\SI{10}{\peta\watt}+$) pulses at high repetition rates, assuming the beam leaks through a mirror after the compressor, this method is suitable to be an \textit{in-situ} diagnostic for the peak intensity. Alternatively, a reflection from a blast-shield glass plate or transmission through a turning mirror are equally well-suited to transport the beam into a secondary chamber to perform ionization measurements there. This allows measuring the peak intensity regardless of the experiment conducted in the main interaction chamber.

To conclude, a new method to diagnose the peak intensity of ultra-intense laser pulses has been presented. It can be easily implemented, requires modest computational resources and access to the modified SMILEI PIC simulation code (that has the three-piece field-ionization rate from \cite{Golovanov2020} included: our addition of this rate is freely available from \cite{iouatu_git}).
The 15,000 1D PIC simulations run on a quad-core laptop in one day of wall clock time, similarly for the integration routines.
We envision that this peak intensity diagnostic will be routinely employed at O(10PW) and 100PW laser facilities.
As the focused intensity is of fundamental importance in all high energy density physics experiments envisaged on these devices, an accurate measurement is required.
This work allows transitioning from currently performed commissioning experiments on O(10PW) laser facilities to the community's first experiments by providing an invaluable tool to assess the true intensity on the target. 

\begin{acknowledgments}
Part of the simulations presented herein were run using the ARCHER2 UK National Supercomputing Service (https://www.archer2.ac.uk) and EPSRC grant number EP/R029148/1. 
The authors gratefully acknowledge the support of the staff of the Central Laser Facility, Rutherford Appleton Laboratory. 
I. Ouatu acknowledges support from an EPSRC DTP grant to Oxford. 
C. Ticos acknowledges support from the PN19060105-2021 grant from MCID.
R. Aboushelbaya and P. A. Norreys acknowledge support from Oxford-ShanghaiTech collaboration agreement. 
K. M. Krushelnick and P. A. Norreys also acknowledge support from the Leverhulme Trust.

\end{acknowledgments}

\bibliographystyle{apsrev4-2}
\bibliography{bibliography_corrections1}

\begin{thebibliography}{33}%
\makeatletter
\providecommand \@ifxundefined [1]{%
 \@ifx{#1\undefined}
}%
\providecommand \@ifnum [1]{%
 \ifnum #1\expandafter \@firstoftwo
 \else \expandafter \@secondoftwo
 \fi
}%
\providecommand \@ifx [1]{%
 \ifx #1\expandafter \@firstoftwo
 \else \expandafter \@secondoftwo
 \fi
}%
\providecommand \natexlab [1]{#1}%
\providecommand \enquote  [1]{``#1''}%
\providecommand \bibnamefont  [1]{#1}%
\providecommand \bibfnamefont [1]{#1}%
\providecommand \citenamefont [1]{#1}%
\providecommand \href@noop [0]{\@secondoftwo}%
\providecommand \href [0]{\begingroup \@sanitize@url \@href}%
\providecommand \@href[1]{\@@startlink{#1}\@@href}%
\providecommand \@@href[1]{\endgroup#1\@@endlink}%
\providecommand \@sanitize@url [0]{\catcode `\\12\catcode `\$12\catcode
  `\&12\catcode `\#12\catcode `\^12\catcode `\_12\catcode `\%12\relax}%
\providecommand \@@startlink[1]{}%
\providecommand \@@endlink[0]{}%
\providecommand \url  [0]{\begingroup\@sanitize@url \@url }%
\providecommand \@url [1]{\endgroup\@href {#1}{\urlprefix }}%
\providecommand \urlprefix  [0]{URL }%
\providecommand \Eprint [0]{\href }%
\providecommand \doibase [0]{https://doi.org/}%
\providecommand \selectlanguage [0]{\@gobble}%
\providecommand \bibinfo  [0]{\@secondoftwo}%
\providecommand \bibfield  [0]{\@secondoftwo}%
\providecommand \translation [1]{[#1]}%
\providecommand \BibitemOpen [0]{}%
\providecommand \bibitemStop [0]{}%
\providecommand \bibitemNoStop [0]{.\EOS\space}%
\providecommand \EOS [0]{\spacefactor3000\relax}%
\providecommand \BibitemShut  [1]{\csname bibitem#1\endcsname}%
\let\auto@bib@innerbib\@empty
\bibitem [{\citenamefont {Danson}\ \emph {et~al.}(19ed)\citenamefont {Danson},
  \citenamefont {Haefner}, \citenamefont {Bromage}, \citenamefont {Butcher},
  \citenamefont {Chanteloup}, \citenamefont {Chowdhury}, \citenamefont
  {Galvanauskas}, \citenamefont {Gizzi}, \citenamefont {Hein}, \citenamefont
  {Hillier}, \citenamefont {Hopps}, \citenamefont {Kato}, \citenamefont
  {Khazanov}, \citenamefont {Kodama}, \citenamefont {Korn}, \citenamefont {Li},
  \citenamefont {Li}, \citenamefont {Limpert}, \citenamefont {Ma},
  \citenamefont {Nam}, \citenamefont {Neely}, \citenamefont {Papadopoulos},
  \citenamefont {Penman}, \citenamefont {Qian}, \citenamefont {Rocca},
  \citenamefont {Shaykin}, \citenamefont {Siders}, \citenamefont {Spindloe},
  \citenamefont {Szatm{\'a}ri}, \citenamefont {Trines}, \citenamefont {Zhu},
  \citenamefont {Zhu},\ and\ \citenamefont
  {Zuegel}}]{ColinDansonFacilitiesReview}%
  \BibitemOpen
  \bibfield  {author} {\bibinfo {author} {\bibfnamefont {C.~N.}\ \bibnamefont
  {Danson}}, \bibinfo {author} {\bibfnamefont {C.}~\bibnamefont {Haefner}},
  \bibinfo {author} {\bibfnamefont {J.}~\bibnamefont {Bromage}}, \bibinfo
  {author} {\bibfnamefont {T.}~\bibnamefont {Butcher}}, \bibinfo {author}
  {\bibfnamefont {J.-C.~F.}\ \bibnamefont {Chanteloup}}, \bibinfo {author}
  {\bibfnamefont {E.~A.}\ \bibnamefont {Chowdhury}}, \bibinfo {author}
  {\bibfnamefont {A.}~\bibnamefont {Galvanauskas}}, \bibinfo {author}
  {\bibfnamefont {L.~A.}\ \bibnamefont {Gizzi}}, \bibinfo {author}
  {\bibfnamefont {J.}~\bibnamefont {Hein}}, \bibinfo {author} {\bibfnamefont
  {D.~I.}\ \bibnamefont {Hillier}}, \bibinfo {author} {\bibfnamefont {N.~W.}\
  \bibnamefont {Hopps}}, \bibinfo {author} {\bibfnamefont {Y.}~\bibnamefont
  {Kato}}, \bibinfo {author} {\bibfnamefont {E.~A.}\ \bibnamefont {Khazanov}},
  \bibinfo {author} {\bibfnamefont {R.}~\bibnamefont {Kodama}}, \bibinfo
  {author} {\bibfnamefont {G.}~\bibnamefont {Korn}}, \bibinfo {author}
  {\bibfnamefont {R.}~\bibnamefont {Li}}, \bibinfo {author} {\bibfnamefont
  {Y.}~\bibnamefont {Li}}, \bibinfo {author} {\bibfnamefont {J.}~\bibnamefont
  {Limpert}}, \bibinfo {author} {\bibfnamefont {J.}~\bibnamefont {Ma}},
  \bibinfo {author} {\bibfnamefont {C.~H.}\ \bibnamefont {Nam}}, \bibinfo
  {author} {\bibfnamefont {D.}~\bibnamefont {Neely}}, \bibinfo {author}
  {\bibfnamefont {D.}~\bibnamefont {Papadopoulos}}, \bibinfo {author}
  {\bibfnamefont {R.~R.}\ \bibnamefont {Penman}}, \bibinfo {author}
  {\bibfnamefont {L.}~\bibnamefont {Qian}}, \bibinfo {author} {\bibfnamefont
  {J.~J.}\ \bibnamefont {Rocca}}, \bibinfo {author} {\bibfnamefont {A.~A.}\
  \bibnamefont {Shaykin}}, \bibinfo {author} {\bibfnamefont {C.~W.}\
  \bibnamefont {Siders}}, \bibinfo {author} {\bibfnamefont {C.}~\bibnamefont
  {Spindloe}}, \bibinfo {author} {\bibfnamefont {S.}~\bibnamefont
  {Szatm{\'a}ri}}, \bibinfo {author} {\bibfnamefont {R.~M. G.~M.}\ \bibnamefont
  {Trines}}, \bibinfo {author} {\bibfnamefont {J.}~\bibnamefont {Zhu}},
  \bibinfo {author} {\bibfnamefont {P.}~\bibnamefont {Zhu}},\ and\ \bibinfo
  {author} {\bibfnamefont {J.~D.}\ \bibnamefont {Zuegel}},\ }\bibfield
  {journal} {\bibinfo  {journal} {High Power Laser Science and Engineering}\
  }\textbf {\bibinfo {volume} {7}},\ \href
  {https://doi.org/10.1017/hpl.2019.36} {10.1017/hpl.2019.36} (\bibinfo {year}
  {2019/ed})\BibitemShut {NoStop}%
\bibitem [{\citenamefont {Strickland}\ and\ \citenamefont
  {Mourou}(1985)}]{CPA_mechanism}%
  \BibitemOpen
  \bibfield  {author} {\bibinfo {author} {\bibfnamefont {D.}~\bibnamefont
  {Strickland}}\ and\ \bibinfo {author} {\bibfnamefont {G.}~\bibnamefont
  {Mourou}},\ }\href {https://doi.org/10.1016/0030-4018(85)90151-8} {\bibfield
  {journal} {\bibinfo  {journal} {Optics Communications}\ }\textbf {\bibinfo
  {volume} {55}},\ \bibinfo {pages} {447} (\bibinfo {year} {1985})}\BibitemShut
  {NoStop}%
\bibitem [{\citenamefont {{J.-P. Chambaret}}\ \emph {et~al.}(2010)\citenamefont
  {{J.-P. Chambaret}}, \citenamefont {{O. Chekhlov}}, \citenamefont {{G.
  Cheriaux}}, \citenamefont {{J. Collier}}, \citenamefont {{R. Dabu}},
  \citenamefont {{P. Dombi}}, \citenamefont {{A. M. Dunne}}, \citenamefont {{K.
  Ertel}}, \citenamefont {{P. Georges}}, \citenamefont {{J. Hebling}},
  \citenamefont {{J. Hein}}, \citenamefont {{C. Hernandez-Gomez}},
  \citenamefont {{C. Hooker}}, \citenamefont {{S. Karsch}}, \citenamefont {{G.
  Korn}}, \citenamefont {{F. Krausz}}, \citenamefont {{C. Le Blanc}},
  \citenamefont {{Zs. Major}}, \citenamefont {{F. Mathieu}}, \citenamefont {{T.
  Metzger}}, \citenamefont {{G. Mourou}}, \citenamefont {{P. Nickles}},
  \citenamefont {{K. Osvay}}, \citenamefont {{B. Rus}}, \citenamefont {{W.
  Sandner}}, \citenamefont {{G. Szab\'o}}, \citenamefont {{D. Ursescu}},\ and\
  \citenamefont {{K. Varj\'u}}}]{ELIpillars1}%
  \BibitemOpen
  \bibfield  {author} {\bibinfo {author} {\bibnamefont {{J.-P. Chambaret}}},
  \bibinfo {author} {\bibnamefont {{O. Chekhlov}}}, \bibinfo {author}
  {\bibnamefont {{G. Cheriaux}}}, \bibinfo {author} {\bibnamefont {{J.
  Collier}}}, \bibinfo {author} {\bibnamefont {{R. Dabu}}}, \bibinfo {author}
  {\bibnamefont {{P. Dombi}}}, \bibinfo {author} {\bibnamefont {{A. M.
  Dunne}}}, \bibinfo {author} {\bibnamefont {{K. Ertel}}}, \bibinfo {author}
  {\bibnamefont {{P. Georges}}}, \bibinfo {author} {\bibnamefont {{J.
  Hebling}}}, \bibinfo {author} {\bibnamefont {{J. Hein}}}, \bibinfo {author}
  {\bibnamefont {{C. Hernandez-Gomez}}}, \bibinfo {author} {\bibnamefont {{C.
  Hooker}}}, \bibinfo {author} {\bibnamefont {{S. Karsch}}}, \bibinfo {author}
  {\bibnamefont {{G. Korn}}}, \bibinfo {author} {\bibnamefont {{F. Krausz}}},
  \bibinfo {author} {\bibnamefont {{C. Le Blanc}}}, \bibinfo {author}
  {\bibnamefont {{Zs. Major}}}, \bibinfo {author} {\bibnamefont {{F.
  Mathieu}}}, \bibinfo {author} {\bibnamefont {{T. Metzger}}}, \bibinfo
  {author} {\bibnamefont {{G. Mourou}}}, \bibinfo {author} {\bibnamefont {{P.
  Nickles}}}, \bibinfo {author} {\bibnamefont {{K. Osvay}}}, \bibinfo {author}
  {\bibnamefont {{B. Rus}}}, \bibinfo {author} {\bibnamefont {{W. Sandner}}},
  \bibinfo {author} {\bibnamefont {{G. Szab\'o}}}, \bibinfo {author}
  {\bibnamefont {{D. Ursescu}}},\ and\ \bibinfo {author} {\bibnamefont {{K.
  Varj\'u}}},\ }in\ \href@noop {} {\emph {\bibinfo {booktitle}
  {Proc.{{SPIE}}}}},\ Vol.\ \bibinfo {volume} {7721}\ (\bibinfo {year}
  {2010})\BibitemShut {NoStop}%
\bibitem [{\citenamefont {Weber}\ \emph {et~al.}(2017)\citenamefont {Weber},
  \citenamefont {Bechet}, \citenamefont {Borneis}, \citenamefont {Brabec},
  \citenamefont {Bu{\v c}ka}, \citenamefont {{Chacon-Golcher}}, \citenamefont
  {Ciappina}, \citenamefont {DeMarco}, \citenamefont {Fajstavr}, \citenamefont
  {Falk}, \citenamefont {Garcia}, \citenamefont {Grosz}, \citenamefont {Gu},
  \citenamefont {Hernandez}, \citenamefont {Holec}, \citenamefont {Jane{\v
  c}ka}, \citenamefont {Janta{\v c}}, \citenamefont {Jirka}, \citenamefont
  {Kadlecova}, \citenamefont {Khikhlukha}, \citenamefont {Klimo}, \citenamefont
  {Korn}, \citenamefont {Kramer}, \citenamefont {Kumar}, \citenamefont
  {Lastovi{\v c}ka}, \citenamefont {Lutoslawski}, \citenamefont {Morejon},
  \citenamefont {Ol{\v s}ovcov{\'a}}, \citenamefont {Rajdl}, \citenamefont
  {Renner}, \citenamefont {Rus}, \citenamefont {Singh}, \citenamefont {{\v
  S}mid}, \citenamefont {Sokol}, \citenamefont {Versaci}, \citenamefont
  {Vr{\'a}na}, \citenamefont {Vranic}, \citenamefont {Vysko{\v c}il},
  \citenamefont {Wolf},\ and\ \citenamefont {Yu}}]{ELIpillars2}%
  \BibitemOpen
  \bibfield  {author} {\bibinfo {author} {\bibfnamefont {S.}~\bibnamefont
  {Weber}}, \bibinfo {author} {\bibfnamefont {S.}~\bibnamefont {Bechet}},
  \bibinfo {author} {\bibfnamefont {S.}~\bibnamefont {Borneis}}, \bibinfo
  {author} {\bibfnamefont {L.}~\bibnamefont {Brabec}}, \bibinfo {author}
  {\bibfnamefont {M.}~\bibnamefont {Bu{\v c}ka}}, \bibinfo {author}
  {\bibfnamefont {E.}~\bibnamefont {{Chacon-Golcher}}}, \bibinfo {author}
  {\bibfnamefont {M.}~\bibnamefont {Ciappina}}, \bibinfo {author}
  {\bibfnamefont {M.}~\bibnamefont {DeMarco}}, \bibinfo {author} {\bibfnamefont
  {A.}~\bibnamefont {Fajstavr}}, \bibinfo {author} {\bibfnamefont
  {K.}~\bibnamefont {Falk}}, \bibinfo {author} {\bibfnamefont {E.-R.}\
  \bibnamefont {Garcia}}, \bibinfo {author} {\bibfnamefont {J.}~\bibnamefont
  {Grosz}}, \bibinfo {author} {\bibfnamefont {Y.-J.}\ \bibnamefont {Gu}},
  \bibinfo {author} {\bibfnamefont {J.-C.}\ \bibnamefont {Hernandez}}, \bibinfo
  {author} {\bibfnamefont {M.}~\bibnamefont {Holec}}, \bibinfo {author}
  {\bibfnamefont {P.}~\bibnamefont {Jane{\v c}ka}}, \bibinfo {author}
  {\bibfnamefont {M.}~\bibnamefont {Janta{\v c}}}, \bibinfo {author}
  {\bibfnamefont {M.}~\bibnamefont {Jirka}}, \bibinfo {author} {\bibfnamefont
  {H.}~\bibnamefont {Kadlecova}}, \bibinfo {author} {\bibfnamefont
  {D.}~\bibnamefont {Khikhlukha}}, \bibinfo {author} {\bibfnamefont
  {O.}~\bibnamefont {Klimo}}, \bibinfo {author} {\bibfnamefont
  {G.}~\bibnamefont {Korn}}, \bibinfo {author} {\bibfnamefont {D.}~\bibnamefont
  {Kramer}}, \bibinfo {author} {\bibfnamefont {D.}~\bibnamefont {Kumar}},
  \bibinfo {author} {\bibfnamefont {T.}~\bibnamefont {Lastovi{\v c}ka}},
  \bibinfo {author} {\bibfnamefont {P.}~\bibnamefont {Lutoslawski}}, \bibinfo
  {author} {\bibfnamefont {L.}~\bibnamefont {Morejon}}, \bibinfo {author}
  {\bibfnamefont {V.}~\bibnamefont {Ol{\v s}ovcov{\'a}}}, \bibinfo {author}
  {\bibfnamefont {M.}~\bibnamefont {Rajdl}}, \bibinfo {author} {\bibfnamefont
  {O.}~\bibnamefont {Renner}}, \bibinfo {author} {\bibfnamefont
  {B.}~\bibnamefont {Rus}}, \bibinfo {author} {\bibfnamefont {S.}~\bibnamefont
  {Singh}}, \bibinfo {author} {\bibfnamefont {M.}~\bibnamefont {{\v S}mid}},
  \bibinfo {author} {\bibfnamefont {M.}~\bibnamefont {Sokol}}, \bibinfo
  {author} {\bibfnamefont {R.}~\bibnamefont {Versaci}}, \bibinfo {author}
  {\bibfnamefont {R.}~\bibnamefont {Vr{\'a}na}}, \bibinfo {author}
  {\bibfnamefont {M.}~\bibnamefont {Vranic}}, \bibinfo {author} {\bibfnamefont
  {J.}~\bibnamefont {Vysko{\v c}il}}, \bibinfo {author} {\bibfnamefont
  {A.}~\bibnamefont {Wolf}},\ and\ \bibinfo {author} {\bibfnamefont
  {Q.}~\bibnamefont {Yu}},\ }\href {https://doi.org/10.1016/j.mre.2017.03.003}
  {\bibfield  {journal} {\bibinfo  {journal} {Matter and Radiation at
  Extremes}\ }\textbf {\bibinfo {volume} {2}},\ \bibinfo {pages} {149}
  (\bibinfo {year} {2017})}\BibitemShut {NoStop}%
\bibitem [{\citenamefont {Lureau}\ \emph {et~al.}(2020)\citenamefont {Lureau},
  \citenamefont {Matras}, \citenamefont {Chalus}, \citenamefont {Derycke},
  \citenamefont {Morbieu}, \citenamefont {Radier}, \citenamefont {Casagrande},
  \citenamefont {Laux}, \citenamefont {Ricaud}, \citenamefont {Rey},
  \citenamefont {Pellegrina}, \citenamefont {Richard}, \citenamefont
  {Boudjemaa}, \citenamefont {{Simon-Boisson}}, \citenamefont {Baleanu},
  \citenamefont {Banici}, \citenamefont {Gradinariu}, \citenamefont
  {Caldararu}, \citenamefont {de~Boisdeffre}, \citenamefont {Ghenuche},
  \citenamefont {Naziru}, \citenamefont {Kolliopoulos}, \citenamefont {Neagu},
  \citenamefont {Dabu}, \citenamefont {Dancus},\ and\ \citenamefont
  {Ursescu}}]{ELINP_progressreport}%
  \BibitemOpen
  \bibfield  {author} {\bibinfo {author} {\bibfnamefont {F.}~\bibnamefont
  {Lureau}}, \bibinfo {author} {\bibfnamefont {G.}~\bibnamefont {Matras}},
  \bibinfo {author} {\bibfnamefont {O.}~\bibnamefont {Chalus}}, \bibinfo
  {author} {\bibfnamefont {C.}~\bibnamefont {Derycke}}, \bibinfo {author}
  {\bibfnamefont {T.}~\bibnamefont {Morbieu}}, \bibinfo {author} {\bibfnamefont
  {C.}~\bibnamefont {Radier}}, \bibinfo {author} {\bibfnamefont
  {O.}~\bibnamefont {Casagrande}}, \bibinfo {author} {\bibfnamefont
  {S.}~\bibnamefont {Laux}}, \bibinfo {author} {\bibfnamefont {S.}~\bibnamefont
  {Ricaud}}, \bibinfo {author} {\bibfnamefont {G.}~\bibnamefont {Rey}},
  \bibinfo {author} {\bibfnamefont {A.}~\bibnamefont {Pellegrina}}, \bibinfo
  {author} {\bibfnamefont {C.}~\bibnamefont {Richard}}, \bibinfo {author}
  {\bibfnamefont {L.}~\bibnamefont {Boudjemaa}}, \bibinfo {author}
  {\bibfnamefont {C.}~\bibnamefont {{Simon-Boisson}}}, \bibinfo {author}
  {\bibfnamefont {A.}~\bibnamefont {Baleanu}}, \bibinfo {author} {\bibfnamefont
  {R.}~\bibnamefont {Banici}}, \bibinfo {author} {\bibfnamefont
  {A.}~\bibnamefont {Gradinariu}}, \bibinfo {author} {\bibfnamefont
  {C.}~\bibnamefont {Caldararu}}, \bibinfo {author} {\bibfnamefont
  {B.}~\bibnamefont {de~Boisdeffre}}, \bibinfo {author} {\bibfnamefont
  {P.}~\bibnamefont {Ghenuche}}, \bibinfo {author} {\bibfnamefont
  {A.}~\bibnamefont {Naziru}}, \bibinfo {author} {\bibfnamefont
  {G.}~\bibnamefont {Kolliopoulos}}, \bibinfo {author} {\bibfnamefont
  {L.}~\bibnamefont {Neagu}}, \bibinfo {author} {\bibfnamefont
  {R.}~\bibnamefont {Dabu}}, \bibinfo {author} {\bibfnamefont {I.}~\bibnamefont
  {Dancus}},\ and\ \bibinfo {author} {\bibfnamefont {D.}~\bibnamefont
  {Ursescu}},\ }\href {https://doi.org/10.1017/hpl.2020.41} {\bibfield
  {journal} {\bibinfo  {journal} {High Power Laser Science and Engineering}\
  }\textbf {\bibinfo {volume} {8}},\ \bibinfo {pages} {e43} (\bibinfo {year}
  {2020})}\BibitemShut {NoStop}%
\bibitem [{UK_()}]{UK_Laser_VULCAN}%
  \BibitemOpen
  \href@noop {} {\bibinfo {title} {{{CLF Vulcan}} 2020 {{Upgrade}}}},\ \bibinfo
  {howpublished}
  {https://www.clf.stfc.ac.uk/Pages/Vulcan-2020.aspx}\BibitemShut {NoStop}%
\bibitem [{\citenamefont {Le~Garrec}\ and\ \citenamefont {{and the Apollon
  team}}(2017)}]{Apollon_France}%
  \BibitemOpen
  \bibfield  {author} {\bibinfo {author} {\bibfnamefont {B.}~\bibnamefont
  {Le~Garrec}}\ and\ \bibinfo {author} {\bibnamefont {{and the Apollon team}}}\
  }(\bibinfo  {publisher} {{Optical Society of America}},\ \bibinfo {year}
  {2017})\ p.\ \bibinfo {pages} {SF1K.3}\BibitemShut {NoStop}%
\bibitem [{\citenamefont {Nees}\ \emph {et~al.}(2020)\citenamefont {Nees},
  \citenamefont {Maksimchuk}, \citenamefont {Kalinchenko}, \citenamefont {Hou},
  \citenamefont {Ma}, \citenamefont {Campbell}, \citenamefont {McKelvey},
  \citenamefont {Willingale}, \citenamefont {Jovanovic}, \citenamefont
  {Kuranz}, \citenamefont {Thomas},\ and\ \citenamefont
  {Krushelnick}}]{ZEUS_Michigan}%
  \BibitemOpen
  \bibfield  {author} {\bibinfo {author} {\bibfnamefont {J.}~\bibnamefont
  {Nees}}, \bibinfo {author} {\bibfnamefont {A.}~\bibnamefont {Maksimchuk}},
  \bibinfo {author} {\bibfnamefont {G.}~\bibnamefont {Kalinchenko}}, \bibinfo
  {author} {\bibfnamefont {B.}~\bibnamefont {Hou}}, \bibinfo {author}
  {\bibfnamefont {Y.}~\bibnamefont {Ma}}, \bibinfo {author} {\bibfnamefont
  {P.}~\bibnamefont {Campbell}}, \bibinfo {author} {\bibfnamefont
  {A.}~\bibnamefont {McKelvey}}, \bibinfo {author} {\bibfnamefont
  {L.}~\bibnamefont {Willingale}}, \bibinfo {author} {\bibfnamefont
  {I.}~\bibnamefont {Jovanovic}}, \bibinfo {author} {\bibfnamefont
  {C.}~\bibnamefont {Kuranz}}, \bibinfo {author} {\bibfnamefont
  {A.}~\bibnamefont {Thomas}},\ and\ \bibinfo {author} {\bibfnamefont
  {K.}~\bibnamefont {Krushelnick}}\ }(\bibinfo  {publisher} {{Optical Society
  of America}},\ \bibinfo {year} {2020})\ p.\ \bibinfo {pages}
  {JW2B.9}\BibitemShut {NoStop}%
\bibitem [{\citenamefont {Bromage}\ \emph {et~al.}(19ed)\citenamefont
  {Bromage}, \citenamefont {Bahk}, \citenamefont {Begishev}, \citenamefont
  {Dorrer}, \citenamefont {Guardalben}, \citenamefont {Hoffman}, \citenamefont
  {Oliver}, \citenamefont {Roides}, \citenamefont {Schiesser}, \citenamefont
  {Iii}, \citenamefont {Spilatro}, \citenamefont {Webb}, \citenamefont
  {Weiner},\ and\ \citenamefont {Zuegel}}]{OPAL_Rochester}%
  \BibitemOpen
  \bibfield  {author} {\bibinfo {author} {\bibfnamefont {J.}~\bibnamefont
  {Bromage}}, \bibinfo {author} {\bibfnamefont {S.-W.}\ \bibnamefont {Bahk}},
  \bibinfo {author} {\bibfnamefont {I.~A.}\ \bibnamefont {Begishev}}, \bibinfo
  {author} {\bibfnamefont {C.}~\bibnamefont {Dorrer}}, \bibinfo {author}
  {\bibfnamefont {M.~J.}\ \bibnamefont {Guardalben}}, \bibinfo {author}
  {\bibfnamefont {B.~N.}\ \bibnamefont {Hoffman}}, \bibinfo {author}
  {\bibfnamefont {J.~B.}\ \bibnamefont {Oliver}}, \bibinfo {author}
  {\bibfnamefont {R.~G.}\ \bibnamefont {Roides}}, \bibinfo {author}
  {\bibfnamefont {E.~M.}\ \bibnamefont {Schiesser}}, \bibinfo {author}
  {\bibfnamefont {M.~J.~S.}\ \bibnamefont {Iii}}, \bibinfo {author}
  {\bibfnamefont {M.}~\bibnamefont {Spilatro}}, \bibinfo {author}
  {\bibfnamefont {B.}~\bibnamefont {Webb}}, \bibinfo {author} {\bibfnamefont
  {D.}~\bibnamefont {Weiner}},\ and\ \bibinfo {author} {\bibfnamefont {J.~D.}\
  \bibnamefont {Zuegel}},\ }\bibfield  {journal} {\bibinfo  {journal} {High
  Power Laser Science and Engineering}\ }\textbf {\bibinfo {volume} {7}},\
  \href {https://doi.org/10.1017/hpl.2018.64} {10.1017/hpl.2018.64} (\bibinfo
  {year} {2019/ed})\BibitemShut {NoStop}%
\bibitem [{\citenamefont {Yoon}\ \emph {et~al.}(2021)\citenamefont {Yoon},
  \citenamefont {Kim}, \citenamefont {Choi}, \citenamefont {Sung},
  \citenamefont {Lee}, \citenamefont {Lee},\ and\ \citenamefont
  {Nam}}]{CoReLS_SKorea}%
  \BibitemOpen
  \bibfield  {author} {\bibinfo {author} {\bibfnamefont {J.~W.}\ \bibnamefont
  {Yoon}}, \bibinfo {author} {\bibfnamefont {Y.~G.}\ \bibnamefont {Kim}},
  \bibinfo {author} {\bibfnamefont {I.~W.}\ \bibnamefont {Choi}}, \bibinfo
  {author} {\bibfnamefont {J.~H.}\ \bibnamefont {Sung}}, \bibinfo {author}
  {\bibfnamefont {H.~W.}\ \bibnamefont {Lee}}, \bibinfo {author} {\bibfnamefont
  {S.~K.}\ \bibnamefont {Lee}},\ and\ \bibinfo {author} {\bibfnamefont {C.~H.}\
  \bibnamefont {Nam}},\ }\href {https://doi.org/10.1364/OPTICA.420520}
  {\bibfield  {journal} {\bibinfo  {journal} {Optica}\ }\textbf {\bibinfo
  {volume} {8}},\ \bibinfo {pages} {630} (\bibinfo {year} {2021})}\BibitemShut
  {NoStop}%
\bibitem [{\citenamefont {Shen}\ \emph {et~al.}(2018)\citenamefont {Shen},
  \citenamefont {Bu}, \citenamefont {Xu}, \citenamefont {Xu}, \citenamefont
  {Ji}, \citenamefont {Li},\ and\ \citenamefont {Xu}}]{SEL_China}%
  \BibitemOpen
  \bibfield  {author} {\bibinfo {author} {\bibfnamefont {B.}~\bibnamefont
  {Shen}}, \bibinfo {author} {\bibfnamefont {Z.}~\bibnamefont {Bu}}, \bibinfo
  {author} {\bibfnamefont {J.}~\bibnamefont {Xu}}, \bibinfo {author}
  {\bibfnamefont {T.}~\bibnamefont {Xu}}, \bibinfo {author} {\bibfnamefont
  {L.}~\bibnamefont {Ji}}, \bibinfo {author} {\bibfnamefont {R.}~\bibnamefont
  {Li}},\ and\ \bibinfo {author} {\bibfnamefont {Z.}~\bibnamefont {Xu}},\
  }\href {https://doi.org/10.1088/1361-6587/aaa7fb} {\bibfield  {journal}
  {\bibinfo  {journal} {Plasma Physics and Controlled Fusion}\ }\textbf
  {\bibinfo {volume} {60}},\ \bibinfo {pages} {044002} (\bibinfo {year}
  {2018})}\BibitemShut {NoStop}%
\bibitem [{\citenamefont {Bashinov}\ \emph {et~al.}(2014)\citenamefont
  {Bashinov}, \citenamefont {Gonoskov}, \citenamefont {Kim}, \citenamefont
  {Mourou},\ and\ \citenamefont {Sergeev}}]{Russia_Laser}%
  \BibitemOpen
  \bibfield  {author} {\bibinfo {author} {\bibfnamefont {A.}~\bibnamefont
  {Bashinov}}, \bibinfo {author} {\bibfnamefont {A.}~\bibnamefont {Gonoskov}},
  \bibinfo {author} {\bibfnamefont {A.}~\bibnamefont {Kim}}, \bibinfo {author}
  {\bibfnamefont {G.}~\bibnamefont {Mourou}},\ and\ \bibinfo {author}
  {\bibfnamefont {A.}~\bibnamefont {Sergeev}},\ }\href
  {https://doi.org/10.1140/epjst/e2014-02161-7} {\bibfield  {journal} {\bibinfo
   {journal} {The European Physical Journal Special Topics}\ }\textbf {\bibinfo
  {volume} {223}},\ \bibinfo {pages} {1105} (\bibinfo {year}
  {2014})}\BibitemShut {NoStop}%
\bibitem [{\citenamefont {Artemenko}\ and\ \citenamefont
  {Kostyukov}(2017)}]{Artmenko2017}%
  \BibitemOpen
  \bibfield  {author} {\bibinfo {author} {\bibfnamefont {I.~I.}\ \bibnamefont
  {Artemenko}}\ and\ \bibinfo {author} {\bibfnamefont {I.~Y.}\ \bibnamefont
  {Kostyukov}},\ }\href {https://doi.org/10.1103/PhysRevA.96.032106} {\bibfield
   {journal} {\bibinfo  {journal} {Physical Review A}\ }\textbf {\bibinfo
  {volume} {96}},\ \bibinfo {pages} {032106} (\bibinfo {year}
  {2017})}\BibitemShut {NoStop}%
\bibitem [{\citenamefont {Bell}\ and\ \citenamefont
  {Kirk}(2008)}]{Bell_Kirk_PRL2008}%
  \BibitemOpen
  \bibfield  {author} {\bibinfo {author} {\bibfnamefont {A.~R.}\ \bibnamefont
  {Bell}}\ and\ \bibinfo {author} {\bibfnamefont {J.~G.}\ \bibnamefont
  {Kirk}},\ }\href {https://doi.org/10.1103/PhysRevLett.101.200403} {\bibfield
  {journal} {\bibinfo  {journal} {Physical Review Letters}\ }\textbf {\bibinfo
  {volume} {101}},\ \bibinfo {pages} {200403} (\bibinfo {year}
  {2008})}\BibitemShut {NoStop}%
\bibitem [{\citenamefont {Ridgers}\ \emph {et~al.}(2012)\citenamefont
  {Ridgers}, \citenamefont {Brady}, \citenamefont {Duclous}, \citenamefont
  {Kirk}, \citenamefont {Bennett}, \citenamefont {Arber}, \citenamefont
  {Robinson},\ and\ \citenamefont {Bell}}]{Bell_Ridgers_PRL2012}%
  \BibitemOpen
  \bibfield  {author} {\bibinfo {author} {\bibfnamefont {C.~P.}\ \bibnamefont
  {Ridgers}}, \bibinfo {author} {\bibfnamefont {C.~S.}\ \bibnamefont {Brady}},
  \bibinfo {author} {\bibfnamefont {R.}~\bibnamefont {Duclous}}, \bibinfo
  {author} {\bibfnamefont {J.~G.}\ \bibnamefont {Kirk}}, \bibinfo {author}
  {\bibfnamefont {K.}~\bibnamefont {Bennett}}, \bibinfo {author} {\bibfnamefont
  {T.~D.}\ \bibnamefont {Arber}}, \bibinfo {author} {\bibfnamefont {A.~P.~L.}\
  \bibnamefont {Robinson}},\ and\ \bibinfo {author} {\bibfnamefont {A.~R.}\
  \bibnamefont {Bell}},\ }\href
  {https://doi.org/10.1103/PhysRevLett.108.165006} {\bibfield  {journal}
  {\bibinfo  {journal} {Physical Review Letters}\ }\textbf {\bibinfo {volume}
  {108}},\ \bibinfo {pages} {165006} (\bibinfo {year} {2012})}\BibitemShut
  {NoStop}%
\bibitem [{\citenamefont {Turcu}\ \emph {et~al.}(2016)\citenamefont {Turcu},
  \citenamefont {Negoita}, \citenamefont {Jaroszynski}, \citenamefont
  {Mckenna}, \citenamefont {Balascuta}, \citenamefont {Ursescu}, \citenamefont
  {Dancus}, \citenamefont {Cernaianu}, \citenamefont {Tataru}, \citenamefont
  {Ghenuche}, \citenamefont {Stutman}, \citenamefont {Boianu}, \citenamefont
  {Risca}, \citenamefont {Toma}, \citenamefont {Petcu}, \citenamefont {Acbas},
  \citenamefont {Yoffe}, \citenamefont {Noble}, \citenamefont {Ersfeld},
  \citenamefont {Brunetti}, \citenamefont {Capdessus}, \citenamefont {Murphy},
  \citenamefont {Ridgers}, \citenamefont {Neely}, \citenamefont {Mangles},
  \citenamefont {Gray}, \citenamefont {Thomas}, \citenamefont {Kirk},
  \citenamefont {Ilderton}, \citenamefont {Marklund}, \citenamefont {Gordon},
  \citenamefont {Hafizi}, \citenamefont {Kaganovich}, \citenamefont {Palastro},
  \citenamefont {D'Humieres}, \citenamefont {Zepf}, \citenamefont {Sarri},
  \citenamefont {Gies}, \citenamefont {Karbstein}, \citenamefont {Schreiber},
  \citenamefont {Paulus}, \citenamefont {Dromey}, \citenamefont {Harvey},
  \citenamefont {Piazza}, \citenamefont {Keitel}, \citenamefont {Kaluza},
  \citenamefont {Gales},\ and\ \citenamefont
  {Zamfir}}]{10PW_enabling_science1}%
  \BibitemOpen
  \bibfield  {author} {\bibinfo {author} {\bibfnamefont {I.~C.~E.}\
  \bibnamefont {Turcu}}, \bibinfo {author} {\bibfnamefont {F.}~\bibnamefont
  {Negoita}}, \bibinfo {author} {\bibfnamefont {D.~A.}\ \bibnamefont
  {Jaroszynski}}, \bibinfo {author} {\bibfnamefont {P.}~\bibnamefont
  {Mckenna}}, \bibinfo {author} {\bibfnamefont {S.}~\bibnamefont {Balascuta}},
  \bibinfo {author} {\bibfnamefont {D.}~\bibnamefont {Ursescu}}, \bibinfo
  {author} {\bibfnamefont {I.}~\bibnamefont {Dancus}}, \bibinfo {author}
  {\bibfnamefont {M.~O.}\ \bibnamefont {Cernaianu}}, \bibinfo {author}
  {\bibfnamefont {M.~V.}\ \bibnamefont {Tataru}}, \bibinfo {author}
  {\bibfnamefont {P.}~\bibnamefont {Ghenuche}}, \bibinfo {author}
  {\bibfnamefont {D.}~\bibnamefont {Stutman}}, \bibinfo {author} {\bibfnamefont
  {A.}~\bibnamefont {Boianu}}, \bibinfo {author} {\bibfnamefont
  {M.}~\bibnamefont {Risca}}, \bibinfo {author} {\bibfnamefont
  {M.}~\bibnamefont {Toma}}, \bibinfo {author} {\bibfnamefont {C.}~\bibnamefont
  {Petcu}}, \bibinfo {author} {\bibfnamefont {G.}~\bibnamefont {Acbas}},
  \bibinfo {author} {\bibfnamefont {S.~R.}\ \bibnamefont {Yoffe}}, \bibinfo
  {author} {\bibfnamefont {A.}~\bibnamefont {Noble}}, \bibinfo {author}
  {\bibfnamefont {B.}~\bibnamefont {Ersfeld}}, \bibinfo {author} {\bibfnamefont
  {E.}~\bibnamefont {Brunetti}}, \bibinfo {author} {\bibfnamefont
  {R.}~\bibnamefont {Capdessus}}, \bibinfo {author} {\bibfnamefont
  {C.}~\bibnamefont {Murphy}}, \bibinfo {author} {\bibfnamefont {C.~P.}\
  \bibnamefont {Ridgers}}, \bibinfo {author} {\bibfnamefont {D.}~\bibnamefont
  {Neely}}, \bibinfo {author} {\bibfnamefont {S.~P.~D.}\ \bibnamefont
  {Mangles}}, \bibinfo {author} {\bibfnamefont {R.~J.}\ \bibnamefont {Gray}},
  \bibinfo {author} {\bibfnamefont {A.~G.~R.}\ \bibnamefont {Thomas}}, \bibinfo
  {author} {\bibfnamefont {J.~G.}\ \bibnamefont {Kirk}}, \bibinfo {author}
  {\bibfnamefont {A.}~\bibnamefont {Ilderton}}, \bibinfo {author}
  {\bibfnamefont {M.}~\bibnamefont {Marklund}}, \bibinfo {author}
  {\bibfnamefont {D.~F.}\ \bibnamefont {Gordon}}, \bibinfo {author}
  {\bibfnamefont {B.}~\bibnamefont {Hafizi}}, \bibinfo {author} {\bibfnamefont
  {D.}~\bibnamefont {Kaganovich}}, \bibinfo {author} {\bibfnamefont {J.~P.}\
  \bibnamefont {Palastro}}, \bibinfo {author} {\bibfnamefont {E.}~\bibnamefont
  {D'Humieres}}, \bibinfo {author} {\bibfnamefont {M.}~\bibnamefont {Zepf}},
  \bibinfo {author} {\bibfnamefont {G.}~\bibnamefont {Sarri}}, \bibinfo
  {author} {\bibfnamefont {H.}~\bibnamefont {Gies}}, \bibinfo {author}
  {\bibfnamefont {F.}~\bibnamefont {Karbstein}}, \bibinfo {author}
  {\bibfnamefont {J.}~\bibnamefont {Schreiber}}, \bibinfo {author}
  {\bibfnamefont {G.~G.}\ \bibnamefont {Paulus}}, \bibinfo {author}
  {\bibfnamefont {B.}~\bibnamefont {Dromey}}, \bibinfo {author} {\bibfnamefont
  {C.}~\bibnamefont {Harvey}}, \bibinfo {author} {\bibfnamefont {A.~D.}\
  \bibnamefont {Piazza}}, \bibinfo {author} {\bibfnamefont {C.~H.}\
  \bibnamefont {Keitel}}, \bibinfo {author} {\bibfnamefont {M.~C.}\
  \bibnamefont {Kaluza}}, \bibinfo {author} {\bibfnamefont {S.}~\bibnamefont
  {Gales}},\ and\ \bibinfo {author} {\bibfnamefont {N.~V.}\ \bibnamefont
  {Zamfir}},\ }\href@noop {} {\bibfield  {journal} {\bibinfo  {journal} {Rom.
  Rep. Phys.}\ ,\ \bibinfo {pages} {87}} (\bibinfo {year} {2016})}\BibitemShut
  {NoStop}%
\bibitem [{\citenamefont {Yanovsky}\ \emph {et~al.}(2008)\citenamefont
  {Yanovsky}, \citenamefont {Chvykov}, \citenamefont {Kalinchenko},
  \citenamefont {Rousseau}, \citenamefont {Planchon}, \citenamefont {Matsuoka},
  \citenamefont {Maksimchuk}, \citenamefont {Nees}, \citenamefont {Cheriaux},
  \citenamefont {Mourou},\ and\ \citenamefont {Krushelnick}}]{LoI_ref1}%
  \BibitemOpen
  \bibfield  {author} {\bibinfo {author} {\bibfnamefont {V.}~\bibnamefont
  {Yanovsky}}, \bibinfo {author} {\bibfnamefont {V.}~\bibnamefont {Chvykov}},
  \bibinfo {author} {\bibfnamefont {G.}~\bibnamefont {Kalinchenko}}, \bibinfo
  {author} {\bibfnamefont {P.}~\bibnamefont {Rousseau}}, \bibinfo {author}
  {\bibfnamefont {T.}~\bibnamefont {Planchon}}, \bibinfo {author}
  {\bibfnamefont {T.}~\bibnamefont {Matsuoka}}, \bibinfo {author}
  {\bibfnamefont {A.}~\bibnamefont {Maksimchuk}}, \bibinfo {author}
  {\bibfnamefont {J.}~\bibnamefont {Nees}}, \bibinfo {author} {\bibfnamefont
  {G.}~\bibnamefont {Cheriaux}}, \bibinfo {author} {\bibfnamefont
  {G.}~\bibnamefont {Mourou}},\ and\ \bibinfo {author} {\bibfnamefont
  {K.}~\bibnamefont {Krushelnick}},\ }\href
  {https://doi.org/10.1364/OE.16.002109} {\bibfield  {journal} {\bibinfo
  {journal} {Optics Express}\ }\textbf {\bibinfo {volume} {16}},\ \bibinfo
  {pages} {2109} (\bibinfo {year} {2008})}\BibitemShut {NoStop}%
\bibitem [{\citenamefont {Pariente}\ \emph {et~al.}(2016)\citenamefont
  {Pariente}, \citenamefont {Gallet}, \citenamefont {Borot}, \citenamefont
  {Gobert},\ and\ \citenamefont {Qu{\'e}r{\'e}}}]{STCs}%
  \BibitemOpen
  \bibfield  {author} {\bibinfo {author} {\bibfnamefont {G.}~\bibnamefont
  {Pariente}}, \bibinfo {author} {\bibfnamefont {V.}~\bibnamefont {Gallet}},
  \bibinfo {author} {\bibfnamefont {A.}~\bibnamefont {Borot}}, \bibinfo
  {author} {\bibfnamefont {O.}~\bibnamefont {Gobert}},\ and\ \bibinfo {author}
  {\bibfnamefont {F.}~\bibnamefont {Qu{\'e}r{\'e}}},\ }\href
  {https://doi.org/10.1038/nphoton.2016.140} {\bibfield  {journal} {\bibinfo
  {journal} {Nature Photonics}\ }\textbf {\bibinfo {volume} {10}},\ \bibinfo
  {pages} {547} (\bibinfo {year} {2016})}\BibitemShut {NoStop}%
\bibitem [{\citenamefont {Derouillat}\ \emph {et~al.}(2018)\citenamefont
  {Derouillat}, \citenamefont {Beck}, \citenamefont {P{\'e}rez}, \citenamefont
  {Vinci}, \citenamefont {Chiaramello}, \citenamefont {Grassi}, \citenamefont
  {Fl{\'e}}, \citenamefont {Bouchard}, \citenamefont {Plotnikov}, \citenamefont
  {Aunai}, \citenamefont {Dargent}, \citenamefont {Riconda},\ and\
  \citenamefont {Grech}}]{SMILEI}%
  \BibitemOpen
  \bibfield  {author} {\bibinfo {author} {\bibfnamefont {J.}~\bibnamefont
  {Derouillat}}, \bibinfo {author} {\bibfnamefont {A.}~\bibnamefont {Beck}},
  \bibinfo {author} {\bibfnamefont {F.}~\bibnamefont {P{\'e}rez}}, \bibinfo
  {author} {\bibfnamefont {T.}~\bibnamefont {Vinci}}, \bibinfo {author}
  {\bibfnamefont {M.}~\bibnamefont {Chiaramello}}, \bibinfo {author}
  {\bibfnamefont {A.}~\bibnamefont {Grassi}}, \bibinfo {author} {\bibfnamefont
  {M.}~\bibnamefont {Fl{\'e}}}, \bibinfo {author} {\bibfnamefont
  {G.}~\bibnamefont {Bouchard}}, \bibinfo {author} {\bibfnamefont
  {I.}~\bibnamefont {Plotnikov}}, \bibinfo {author} {\bibfnamefont
  {N.}~\bibnamefont {Aunai}}, \bibinfo {author} {\bibfnamefont
  {J.}~\bibnamefont {Dargent}}, \bibinfo {author} {\bibfnamefont
  {C.}~\bibnamefont {Riconda}},\ and\ \bibinfo {author} {\bibfnamefont
  {M.}~\bibnamefont {Grech}},\ }\href
  {https://doi.org/10.1016/j.cpc.2017.09.024} {\bibfield  {journal} {\bibinfo
  {journal} {Computer Physics Communications}\ }\textbf {\bibinfo {volume}
  {222}},\ \bibinfo {pages} {351} (\bibinfo {year} {2018})}\BibitemShut
  {NoStop}%
\bibitem [{\citenamefont {Kostyukov}\ and\ \citenamefont
  {Golovanov}(2018)}]{Kostyukov_Golovanov_2018}%
  \BibitemOpen
  \bibfield  {author} {\bibinfo {author} {\bibfnamefont {I.~Y.}\ \bibnamefont
  {Kostyukov}}\ and\ \bibinfo {author} {\bibfnamefont {A.~A.}\ \bibnamefont
  {Golovanov}},\ }\href {https://doi.org/10.1103/PhysRevA.98.043407} {\bibfield
   {journal} {\bibinfo  {journal} {Physical Review A}\ }\textbf {\bibinfo
  {volume} {98}},\ \bibinfo {pages} {043407} (\bibinfo {year} {2018})},\
  \Eprint {https://arxiv.org/abs/1808.06890} {arXiv:1808.06890} \BibitemShut
  {NoStop}%
\bibitem [{\citenamefont {Kostyukov}\ and\ \citenamefont
  {Golovanov}(2019)}]{Golovanov2020}%
  \BibitemOpen
  \bibfield  {author} {\bibinfo {author} {\bibfnamefont {I.~Y.}\ \bibnamefont
  {Kostyukov}}\ and\ \bibinfo {author} {\bibfnamefont {A.~A.}\ \bibnamefont
  {Golovanov}},\ }\href@noop {} {\bibinfo {title} {Field ionization rate for
  {{PIC}} codes}} (\bibinfo {year} {2019}),\ \Eprint
  {https://arxiv.org/abs/1906.01358} {arXiv:1906.01358 [physics.plasm-ph]}
  \BibitemShut {NoStop}%
\bibitem [{\citenamefont {Bauer}\ and\ \citenamefont {Mulser}(1999)}]{BM1990}%
  \BibitemOpen
  \bibfield  {author} {\bibinfo {author} {\bibfnamefont {D.}~\bibnamefont
  {Bauer}}\ and\ \bibinfo {author} {\bibfnamefont {P.}~\bibnamefont {Mulser}},\
  }\href {https://doi.org/10.1103/PhysRevA.59.569} {\bibfield  {journal}
  {\bibinfo  {journal} {Physical Review A}\ }\textbf {\bibinfo {volume} {59}},\
  \bibinfo {pages} {569} (\bibinfo {year} {1999})}\BibitemShut {NoStop}%
\bibitem [{\citenamefont {Ciappina}\ \emph {et~al.}(2019)\citenamefont
  {Ciappina}, \citenamefont {Popruzhenko}, \citenamefont {Bulanov},
  \citenamefont {Ditmire}, \citenamefont {Korn},\ and\ \citenamefont
  {Weber}}]{Ciappina2019}%
  \BibitemOpen
  \bibfield  {author} {\bibinfo {author} {\bibfnamefont {M.~F.}\ \bibnamefont
  {Ciappina}}, \bibinfo {author} {\bibfnamefont {S.~V.}\ \bibnamefont
  {Popruzhenko}}, \bibinfo {author} {\bibfnamefont {S.~V.}\ \bibnamefont
  {Bulanov}}, \bibinfo {author} {\bibfnamefont {T.}~\bibnamefont {Ditmire}},
  \bibinfo {author} {\bibfnamefont {G.}~\bibnamefont {Korn}},\ and\ \bibinfo
  {author} {\bibfnamefont {S.}~\bibnamefont {Weber}},\ }\href
  {https://doi.org/10.1103/PhysRevA.99.043405} {\bibfield  {journal} {\bibinfo
  {journal} {Physical Review A}\ }\textbf {\bibinfo {volume} {99}},\ \bibinfo
  {pages} {043405} (\bibinfo {year} {2019})}\BibitemShut {NoStop}%
\bibitem [{\citenamefont {Ciappina}\ and\ \citenamefont
  {Popruzhenko}(2020)}]{Ciappina2020_1}%
  \BibitemOpen
  \bibfield  {author} {\bibinfo {author} {\bibfnamefont {M.~F.}\ \bibnamefont
  {Ciappina}}\ and\ \bibinfo {author} {\bibfnamefont {S.~V.}\ \bibnamefont
  {Popruzhenko}},\ }\href {https://doi.org/10.1088/1612-202X/ab6559} {\bibfield
   {journal} {\bibinfo  {journal} {Laser Physics Letters}\ }\textbf {\bibinfo
  {volume} {17}},\ \bibinfo {pages} {025301} (\bibinfo {year}
  {2020})}\BibitemShut {NoStop}%
\bibitem [{\citenamefont {Ciappina}\ \emph {et~al.}(2020)\citenamefont
  {Ciappina}, \citenamefont {Peganov},\ and\ \citenamefont
  {Popruzhenko}}]{Ciappina2020_2}%
  \BibitemOpen
  \bibfield  {author} {\bibinfo {author} {\bibfnamefont {M.~F.}\ \bibnamefont
  {Ciappina}}, \bibinfo {author} {\bibfnamefont {E.~E.}\ \bibnamefont
  {Peganov}},\ and\ \bibinfo {author} {\bibfnamefont {S.~V.}\ \bibnamefont
  {Popruzhenko}},\ }\href {https://doi.org/10.1063/5.0005380} {\bibfield
  {journal} {\bibinfo  {journal} {Matter and Radiation at Extremes}\ }\textbf
  {\bibinfo {volume} {5}},\ \bibinfo {pages} {044401} (\bibinfo {year}
  {2020})}\BibitemShut {NoStop}%
\bibitem [{\citenamefont {Keldysh}(1965)}]{Keldysh1965}%
  \BibitemOpen
  \bibfield  {author} {\bibinfo {author} {\bibfnamefont {L.~V.}\ \bibnamefont
  {Keldysh}},\ }\href@noop {} {\bibfield  {journal} {\bibinfo  {journal} {J.
  Exp. Theor. Phys.}\ }\textbf {\bibinfo {volume} {20}},\ \bibinfo {pages}
  {1307} (\bibinfo {year} {1965})}\BibitemShut {NoStop}%
\bibitem [{\citenamefont {Popov}(2004)}]{Popov2004}%
  \BibitemOpen
  \bibfield  {author} {\bibinfo {author} {\bibfnamefont {V.~S.}\ \bibnamefont
  {Popov}},\ }\href {https://doi.org/10.1070/PU2004v047n09ABEH001812}
  {\bibfield  {journal} {\bibinfo  {journal} {Physics-Uspekhi}\ }\textbf
  {\bibinfo {volume} {47}},\ \bibinfo {pages} {855} (\bibinfo {year}
  {2004})}\BibitemShut {NoStop}%
\bibitem [{\citenamefont {Perelomov}\ \emph {et~al.}(1966)\citenamefont
  {Perelomov}, \citenamefont {Popov},\ and\ \citenamefont
  {Terent'Ev}}]{PPT1966}%
  \BibitemOpen
  \bibfield  {author} {\bibinfo {author} {\bibfnamefont {A.~M.}\ \bibnamefont
  {Perelomov}}, \bibinfo {author} {\bibfnamefont {V.~S.}\ \bibnamefont
  {Popov}},\ and\ \bibinfo {author} {\bibfnamefont {M.~V.}\ \bibnamefont
  {Terent'Ev}},\ }\href@noop {} {\bibfield  {journal} {\bibinfo  {journal}
  {JETP}\ }\textbf {\bibinfo {volume} {23}} (\bibinfo {year}
  {1966})}\BibitemShut {NoStop}%
\bibitem [{\citenamefont {Ammosov}\ and\ \citenamefont
  {Krainov}(1986)}]{ADK1986}%
  \BibitemOpen
  \bibfield  {author} {\bibinfo {author} {\bibfnamefont {M.~V.}\ \bibnamefont
  {Ammosov}}\ and\ \bibinfo {author} {\bibfnamefont {V.~P.}\ \bibnamefont
  {Krainov}},\ }\href@noop {} {\bibfield  {journal} {\bibinfo  {journal} {Zh.
  Eksp. Teor. Fiz.}\ }\textbf {\bibinfo {volume} {91}} (\bibinfo {year}
  {1986})}\BibitemShut {NoStop}%
\bibitem [{\citenamefont {Augst}\ \emph {et~al.}(1991)\citenamefont {Augst},
  \citenamefont {Meyerhofer}, \citenamefont {Strickland},\ and\ \citenamefont
  {Chin}}]{Donna_BSIdiscovery_1990}%
  \BibitemOpen
  \bibfield  {author} {\bibinfo {author} {\bibfnamefont {S.}~\bibnamefont
  {Augst}}, \bibinfo {author} {\bibfnamefont {D.~D.}\ \bibnamefont
  {Meyerhofer}}, \bibinfo {author} {\bibfnamefont {D.}~\bibnamefont
  {Strickland}},\ and\ \bibinfo {author} {\bibfnamefont {S.~L.}\ \bibnamefont
  {Chin}},\ }\href {https://doi.org/10.1364/JOSAB.8.000858} {\bibfield
  {journal} {\bibinfo  {journal} {JOSA B}\ }\textbf {\bibinfo {volume} {8}},\
  \bibinfo {pages} {858} (\bibinfo {year} {1991})}\BibitemShut {NoStop}%
\bibitem [{\citenamefont {Tong}\ and\ \citenamefont
  {Lin}(2005)}]{Tong_Lin_2005}%
  \BibitemOpen
  \bibfield  {author} {\bibinfo {author} {\bibfnamefont {X.~M.}\ \bibnamefont
  {Tong}}\ and\ \bibinfo {author} {\bibfnamefont {C.~D.}\ \bibnamefont {Lin}},\
  }\href {https://doi.org/10.1088/0953-4075/38/15/001} {\bibfield  {journal}
  {\bibinfo  {journal} {Journal of Physics B: Atomic, Molecular and Optical
  Physics}\ }\textbf {\bibinfo {volume} {38}},\ \bibinfo {pages} {2593}
  (\bibinfo {year} {2005})}\BibitemShut {NoStop}%
\bibitem [{\citenamefont {Han}()}]{ElemNumAnalysisIowa}%
  \BibitemOpen
  \bibfield  {author} {\bibinfo {author} {\bibfnamefont {W.}~\bibnamefont
  {Han}},\ }\href@noop {} {\bibinfo {title} {Elementary {{Numerical Analysis
  Course}}}},\ \bibinfo {howpublished}
  {https://homepage.math.uiowa.edu/\textasciitilde
  whan/3800.d/3800.html}\BibitemShut {NoStop}%
\bibitem [{iou()}]{iouatu_git}%
  \BibitemOpen
  \href@noop {} {\bibinfo {title}
  {{{https://github.com/iouatu/mySmilei}}}}\BibitemShut {NoStop}%
\end{thebibliography}%

\end{document}